\documentclass[aps,prb,amsmath,amssymb,reprint,superscriptaddress,showpacs]{revtex4-1}
\usepackage{graphicx}
\usepackage[dvipdfm,colorlinks=true,linkcolor=blue,filecolor=blue,urlcolor=blue]{hyperref}

\usepackage{color}

\begin{document}

\title{Quantum criticality in an asymmetric
three-leg spin tube: A strong rung-coupling perspective}

\author{Yohei Fuji}
\email{fuji@issp.u-tokyo.ac.jp}
\affiliation{Institute for Solid State Physics, University of Tokyo, Kashiwa 277-8581, Japan}

\author{Satoshi Nishimoto}
\affiliation{Institute for Theoretical Solid State Physics, IFW Dresden, D-01171 Dresden, Germany}

\author{Hitoshi Nakada}
\affiliation{Department of Physics, Graduate School of Science, Chiba University, Chiba 263-8522, Japan}

\author{Masaki Oshikawa}
\affiliation{Institute for Solid State Physics, University of Tokyo, Kashiwa 277-8581, Japan}

\date{\today}

\begin{abstract}
We study quantum phase transitions in the
asymmetric variation of the three-leg Heisenberg tube for half-odd-integer spin,
with a modulation of one of the rung exchange couplings $J'_\perp$
while the other two are kept constant $J_\perp$.
We focus on the strong rung-coupling regime $J_\perp \gg J_\parallel$,
where $J_\parallel$ is the leg coupling, and analyze
the effective spin-orbital model with a transverse crystal field in detail.
Applying the Abelian bosonization to the effective model, we find that
the system is in the dimer phase for the general half-odd-integer-spin cases
without the rung modulation; 
the phase transition between the dimer and Tomonaga-Luttinger-liquid 
phases induced by the rung modulation is
of the SU(2)-symmetric Berezinskii-Kosterlitz-Thouless type.
Moreover, we perform a level spectroscopy analysis for the effective
model for spin-1/2 using exact diagonalization, to determine the precise
transition point $| J'_\perp - J_\perp| /J_\parallel \sim 0.283$ in the strong rung-coupling limit.
The presence of the dimer phase in a small but finite region
is also confirmed by a density-matrix renormalization group
calculation on the original spin-tube model.
\end{abstract}


\maketitle

%%%%%%%%%%%%%%%%%%%%%%%%%%%%%%%%%%%%%%%%%%%%%%%%%%%%%%%%%%%%%%%%%%%%

\section{Introduction}
\label{Introduction}

%---[General introduction]----------------------------------------

The quantum Heisenberg antiferromagnet is
one of the most fundamental problems in quantum magnetism.
Quantum fluctuations are stronger in lower dimensions, opening the
possibility of the destruction of the long-range magnetic order.
In fact, the one-dimensional (1D) Heisenberg antiferromagnetic chain
has no long-range order.~\cite{GNT, Giamarchi}
On the other hand, the quantum Heisenberg
antiferromagnet on the two-dimensional (2D) square lattice
does exhibit long-range antiferromagnetic order, even in the
case with spin-$1/2$ where the quantum fluctuations are strongest.
As an intermediate situation between 1D and 2D, a finite
number of coupled chains such as ladder and tube systems may hold
a great potential of providing fascinating new physics from the point of
view of the dimensional crossover.

%---[Haldane conjecture on the spin chain and ladders]-----------------

For the 1D antiferromagnetic Heisenberg chain, the long-range order is
absent for any spin quantum number. However, there is an
important distinction first noticed by Haldane:~\cite{Haldane83a} the
ground state is in a gapless critical phase for half-odd-integer
spins, while it is in a gapped and disordered phase for integer
spins. This classification is later generalized  for antiferromagnetic
quantum spin ladder systems:~\cite{Sierra96, DellAringa97}
denoting the magnitude of the intrinsic spin on each site as $s$ and
the number of chains as $N$, the ladder with $Ns=$ half-odd-integer
belongs to the same universality class as the half-odd-integer spin
chain, while the ladder with $Ns=$ integer has a gapped ground
state. Although these predictions are initially based on the
semi-classical analysis, the validity of them has so far been
confirmed by a significant amount of independent numerical and
analytical calculations.~\cite{White94,Reigrotzki94,Hatano95,Greven96}

%---[Preceding studies about 3-leg spin tubes]--------------------

On the other hand, the physics of
the antiferromagnetic spin ladder is possibly changed if we impose the periodic 
boundary conditions (PBC) in the rung direction, forming a \emph{spin tube}.
In particular, odd-leg spin tubes do not have the
simple N\'eel state as a classical ground state.
The odd-leg spin tubes can be regarded as a simple realization
of \emph{geometrical frustration}; this is one of the reasons
why the spin tubes are of current interest.

A minimal model is the three-leg ``equilateral'' antiferromagnetic $s=1/2$ spin tube, 
in which all the rung exchange interactions are the same.
This model has been studied with a variety of theoretical techniques. 
Strong rung-coupling expansion studies~\cite{Totsuka96, Schulz96, Kawano97, 
Cabra97, Cabra98b, Orignac00, Luscher04} have shown that
the ground state is in a gapped dimer phase, where the spins form singlet pairs 
in alternate shifts and the translational symmetry is broken spontaneously.
This dimer phase is extended beyond the strong rung-coupling region, 
as demonstrated by numerical studies using density-matrix renormalization group 
(DMRG) method.~\cite{Nishimoto08, Sakai08, Sakai10} 

%---[3-leg tube Hamiltonian]------------------------------------

\begin{figure}
\includegraphics[clip,width=6cm]{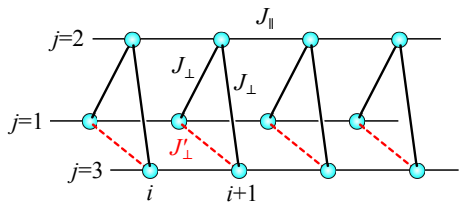}
\caption{(Color online) Schematic picture of the three-leg tube Hamiltonian \eqref{eq:3LegTubeHam}.}
\label{fig:Model}
\end{figure}

The next interesting question is the connection between
the spin tube and the corresponding three-leg spin ladder.
The two models are connected by modulating
rung interactions on one of the edges of the triangular cross section.
The generalized spin tube model,~\cite{Sakai05, Nishimoto08, Sakai08, Sakai10, Charrier10} 
which includes the ``equilateral'' spin tube and the spin ladder
as special cases, is given by the Hamiltonian (see Fig.~\ref{fig:Model}), 
\begin{eqnarray}
H &=& J_\parallel \sum_{i=1}^L \sum_{j=1}^3 {\vec s}_{i,j} \cdot {\vec s}_{i+1,j} \nonumber \\
      & & + J_\perp \sum_{i=1}^L \left( {\vec s}_{i,1} \cdot {\vec s}_{i,2} + {\vec s}_{i,2} \cdot {\vec s}_{i,3} \right) \nonumber \\ 
      & & +J'_\perp \sum_{i=1}^L {\vec s}_{i,1} \cdot {\vec s}_{i,3},
\label{eq:3LegTubeHam}
\end{eqnarray}
where ${\vec s}_{i,j}$ is the spin-$s$ operator, $i$ ($j$) is a rung (leg) index, 
and $L$ is the system length in the leg direction. 
There are three kinds of the exchange couplings: 
$J_\parallel$ for the leg, and $J_\perp$, $J'_\perp$ for the rung. 
In this paper, we assume that all the couplings are antiferromagnetic 
($J_\parallel$, $J_\perp$, $J'_\perp \geq 0$). 
For convenience, the modulation strength is controlled by 
varying $J'_\perp$ with fixed $J_\perp$. 
Hereafter, we call the $J'_\perp = J_\perp$ (equilateral) case as a symmetric
tube and the $J'_\perp \neq J_\perp$ case as an asymmetric tube. 

%---[Our problem]---------------------------------------------------

Let us now focus on the case of $s=1/2$ and $J_\perp \gg J_\parallel$. 
It is established that the symmetric tube is 
in the dimer phase and has a finite excitation gap. 
On the other hand, the system is equivalent to a three-leg ladder 
in the asymmetric limit $J'_\perp \rightarrow 0$, and is divided to 
a single chain and a two-leg ladder in the opposite asymmetric limit 
$J'_\perp \rightarrow \infty$; 
they are both in a gapless critical phase.
Therefore, a phase transition between the gapped and gapless phases
is naturally expected at some value of $|J'_\perp - J_\perp| > 0$.

Surprisingly, it was found~\cite{Sakai05,Nishimoto08} that the gap
quickly vanishes once the asymmetry is introduced by changing
$J'_\perp$ away from $J_\perp$. 
This is the case in particular in the strong rung-coupling limit 
$J_\perp \gg J_\parallel$.
In Ref.~\onlinecite{Nishimoto08}, a DMRG calculation revealed that,
the gap already collapses with the asymmetry of just one
percent $|J'_\perp - J_\perp|/J_\perp \sim 1/100$ at $J_\parallel/J_\perp = 1/100$.
It appears from the numerical result as if the gap is
non-zero only at the symmetric point $J'_\perp = J_\perp$.

However, the conventional wisdom is that any gapped phase extends to 
a finite range of parameter in any local Hamiltonian.  
In fact, in Ref.~\onlinecite{Sakai08} non-Abelian bosonization was applied to 
the three-leg Hubbard model in the `band representation', and it was
suggested that the dimer phase remains in a finite range of the
asymmetric modulation $|J'_\perp - J_\perp| > 0$ and the transition
is of the Berezinskii-Kosterlitz-Thouless (BKT) type. 
They confirmed those theoretical predictions by numerical analyses 
using DMRG and exact diagonalization, and they also found that 
the BKT transition is similar to that in the $s=1/2$ $J_1$-$J_2$ 
chain.~\cite{Haldane82a,*Haldane82b, Okamoto92}
However, analyzing the effect of the small asymmetric modulation is
difficult especially for numerical study.
As a consequence, the quantum phase transition between the gapped 
dimer phase and the gapless phase in the $s=1/2$ three-leg spin
tube has not been quantitatively understood.
The same system with half-odd-integer $s>1/2$ is even less understood.
We note that there are also several studies~\cite{Charrier10,WLi13}
on the three-leg spin tube with integer spin $s$.
However, in this paper we focus on half-odd-integer $s$ including the simplest case $s=1/2$.

As an alternative to the analysis of the original spin
tube model, an effective Hamiltonian in the strong
rung-coupling limit was proposed in Ref.~\onlinecite{Nishimoto08}.
This is indeed useful in understanding why the gap
vanishes at a small but finite asymmetry.~\cite{Sakai08}
However, the effective model has not been studied in detail.
In this paper, in order to clarify the nature of phase transitions
induced by the rung-coupling modulation in the three-leg spin tube~\eqref{eq:3LegTubeHam} 
with half-odd-integer spins, we analyze the effective Hamiltonian
by field-theoretical and numerical methods.

%---[Organization of this paper]------------------------------------------

The paper is organized as follows. 
In Sec.~\ref{SCHam}, we review the derivation of 
the strong rung-coupling effective Hamiltonian for the $s=1/2$ case 
and generalize it to arbitrary half-odd-integer spin cases. 
In Sec.~\ref{FieldTheory}, the low-energy properties of 
the effective Hamiltonian are investigated by Abelian bosonization 
and the renormalization group (RG) analysis. 
In Sec.~\ref{Numerical}, the analytical results are confirmed by 
the level spectroscopy method using exact diagonalization applied to 
the effective Hamiltonian, and the central charge analysis 
for the original spin tube using the DMRG method.
Finally, we summarize our results in Sec.~\ref{Conclusion}.

%%%%%%%%%%%%%%%%%%%%%%%%%%%%%%%%%%%%%%%%%%%%%%%%%%%%%%%%%%%%%%%%%%%%%

\section{Effective Hamiltonian}

In this section, we review the derivation~\cite{Nishimoto08} of
the strong rung-coupling effective Hamiltonian for the asymmetric tube 
by using degenerate perturbation theory ($J_\perp, J'_\perp \gg J_\parallel$). 
The unperturbed part is the sum of decoupled triangles,
\begin{eqnarray}
H_0 &=& \sum_i H_{0,i},\\ H_{0,i} &=& J_\perp \left( {\vec s}_{i,1}
\cdot {\vec s}_{i,2} + {\vec s}_{i,2} \cdot {\vec s}_{i,3} \right) +
J'_\perp {\vec s}_{i,1} \cdot {\vec s}_{i,3}.
\label{eq:3SpinHam}
\end{eqnarray}
We denote the total spin of each triangle, which contains three
spin-$s$ sites, as ${\vec S}_i={\vec s}_{i,1}+{\vec s}_{i,2}+{\vec s}_{i,3}$, 
and the eigenvalues of ${\vec S}_i$, $S^z_i$, and $s_{i,j}^z$ respectively 
as $S$, $M$, and $m_j$, where the rung index $i$ is omitted for brevity.
For each triangle, basis states diagonalizing $s_{i,j}^z$ are denoted 
as $\left| m_1 m_2 m_3 \right>$. 
For later convenience, we also introduce the parity $P = \pm 1$ 
with respect to the exchange of ${\vec s}_{i,1}$ and ${\vec s}_{i,3}$ 
in each triangle.

%---[Review of strong-coupling expansion for the S=1/2 tube]-----------------------

\subsection{Spin-\texorpdfstring{$1/2$}{1/2} case}
\label{SCHam}

First, we consider the case of $s=1/2$ ($m_j = \uparrow, \downarrow$). 
This case has been studied for both the symmetric and
asymmetric tubes by several groups.~\cite{Totsuka96,Schulz96,Nishimoto08,Sakai08}

%---[3-spin problem]---------------------------------------------------

Let us begin by analyzing the decoupled triangle~\eqref{eq:3SpinHam}, 
which has \emph{one} quartet and \emph{two} doublets. 
Energy of the quartet is $E=\frac{J_\perp}{2}+\frac{J'_\perp}{4}$ and 
the eigenstates are written as
\begin{eqnarray}
& \left| \uparrow \uparrow \uparrow \right>, & \nonumber \\
& \frac{1}{\sqrt 3} \left( \left| \uparrow \uparrow \downarrow \right> + \left| \uparrow \downarrow \uparrow \right>
+ \left| \downarrow \uparrow \uparrow \right> \right), & \nonumber \\
& \frac{1}{\sqrt 3} \left( \left| \downarrow \downarrow \uparrow \right> + \left| \downarrow \uparrow \downarrow \right>
+ \left| \uparrow \downarrow \downarrow \right> \right), & \nonumber \\
& \left| \downarrow \downarrow \downarrow \right>. &
\end{eqnarray}
These states have $S=3/2$ and $P=+1$. 
Since we focus on the antiferromagnetic couplings in this paper, 
these higher energy states are neglected in the strong rung-coupling limit. 
The remaining $S=1/2$ doublets are given by
\begin{eqnarray}
& \frac{1}{\sqrt 6} \left( \left| \uparrow \uparrow \downarrow \right> -2 \left| \uparrow \downarrow \uparrow \right>
+ \left| \downarrow \uparrow \uparrow \right> \right) \equiv \left| \uparrow + \right>, & \nonumber \\
& \frac{1}{\sqrt 6} \left( \left| \downarrow \downarrow \uparrow \right> -2 \left| \downarrow \uparrow \downarrow \right>
+ \left| \uparrow \downarrow \downarrow \right> \right) \equiv \left| \downarrow + \right>, &
\end{eqnarray}
with $E=-J_\perp+\frac{J'_\perp}{4}$ and $P=+1$, and 
\begin{eqnarray}
& \frac{1}{\sqrt 2} \left( \left| \uparrow \uparrow \downarrow \right> - \left| \downarrow \uparrow \uparrow \right> \right)
\equiv \left| \uparrow - \right>, & \nonumber \\
& \frac{1}{\sqrt 2} \left( \left| \downarrow \downarrow \uparrow \right> - \left| \uparrow \downarrow \downarrow \right> \right)
\equiv \left| \downarrow - \right>, &
\end{eqnarray}
with $E=-\frac{3J'_\perp}{4}$ and $P=-1$. 
These four states are adopted as the unperturbed states. 
Hereafter, we denote the doublet states as $\left| M P \right>$, 
where $M=\uparrow,\downarrow$ and $P=+,-$.

%---[Strong-coupling Hamiltonian for the S=1/2 tube]--------------------------

By projecting Eq.~\eqref{eq:3LegTubeHam} onto the $4^L$-dimensional Hilbert space 
consisting of the direct products of $\left| M P \right>$'s, 
the effective Hamiltonian is obtained as
\begin{eqnarray}
H &=& \frac{J_\parallel}{3} \sum_{i=1}^L {\vec S}_i \cdot {\vec S}_{i+1} 
          \left[ 1 +8 \left( \tau^x_i \tau^x_{i+1} + \tau^z_i \tau^z_{i+1} \right) \right] \nonumber \\
   & & + \left( J'_\perp - J_\perp \right) \sum_{i=1}^L \tau^z_i,
\label{eq:SCHam}
\end{eqnarray}
up to first order of $J_\parallel$, 
where ${\vec S}_i$ is the spin-$1/2$ operator at site $i$.
The effective model is defined on a 1D chain, 
where each site $i$ represents three sites on the same rung 
in the original tube model.
Here, the parity operators $\tau^{x,y,z}$ are introduced as
\begin{eqnarray}
\tau^z = \frac{1}{2} \left( \left| + \right> \left< + \right| - \left| - \right> \left< - \right| \right), 
\label{eq:TauZ}
\end{eqnarray}
and
\begin{eqnarray}
\tau^x \equiv \frac{1}{2} \left( \tau^+ + \tau^- \right), \hspace{10pt} 
\tau^y \equiv \frac{1}{2i} \left( \tau^+ - \tau^- \right),
\label{eq:Taux}
\end{eqnarray}
with
\begin{eqnarray}
\tau^+ = \left| + \right> \left< - \right|, \hspace{10pt} \tau^- = \left| - \right> \left< + \right|, 
\label{eq:TauXY}
\end{eqnarray}
so that their eigenvalues become $\pm 1/2$ (a half of $P$) as usual spin-$1/2$ operators. 
By applying an appropriate unitary transformation, 
the Hamiltonian~\eqref{eq:SCHam} can be rewritten as
\begin{eqnarray}
H &=& \frac{J_\parallel}{3} \sum_{i=1}^L {\vec S}_i \cdot {\vec S}_{i+1} 
          \left[ 1 +8 \left( \tau^x_i \tau^x_{i+1} + \tau^y_i \tau^y_{i+1} \right) \right] \nonumber \\
   & & + \left( J'_\perp - J_\perp \right) \sum_{i=1}^L \tau^x_i.
\label{eq:SCHam2}
\end{eqnarray}
The last term works like a transverse crystal field and its
coefficient corresponds to the energy difference between the odd- and
even-parity states of the doublets in the unperturbed part.

%---[Preceding studies on the strong-coupling Hamiltonian]

For the symmetric case ($J'_\perp = J_\perp$), the last term of
Eq.~\eqref{eq:SCHam2} vanishes and the effective Hamiltonian 
is simplified as obtained in the preceding studies.~\cite{Totsuka96, Schulz96} 
We thus can regard $\tau$ as an operator acting on \emph{chirality}
corresponding to momenta $\pm 2 \pi /3$ in the rung triangle, 
and spin and chirality are coupled by the leg exchange interaction 
in a biquadratic form. 
It has been confirmed that the system is in a spontaneously 
dimerized state where the spin and chirality form
singlet pairs alternately and have a finite excitation gap.~\cite{Kawano97}

For the asymmetric case ($J'_\perp \neq J_\perp$),
while the effective model~\eqref{eq:SCHam2} was derived 
in Ref.~\onlinecite{Nishimoto08}, no definitive study has been so far reported.
We here start from trivial limits.
If the transverse field is sufficiently strong, the chirality
degrees of freedom are fully polarized, and the Hamiltonian~\eqref{eq:SCHam2} 
is reduced to the spin-$1/2$ Heisenberg chain. 
Hence, the system goes to a gapless critical phase.
In the original spin tube~\eqref{eq:3LegTubeHam} with $s=1/2$, 
these limits correspond to a three-leg open ladder 
($J'_\perp - J_\perp \rightarrow - J_\perp$) and 
a decoupled system consisting of a single chain and a two-leg ladder 
($J'_\perp - J_\perp \rightarrow \infty$), respectively.
In both of these limits, the original spin tube model is gapless.
Thus the effective model correctly describes the origninal spin tube in these limits.
We can further expect that the phase transitions
between the dimer and critical phases are also described by the
effective Hamiltonian~\eqref{eq:SCHam2}.
It is the purpose of the present paper to elucidate the
phase transitions in detail, based on the effective Hamiltonian.

%---[Strong-coupling Hamiltonian for the general-S case]--------------------------

\subsection{General-\texorpdfstring{$s$}{s} case}

Next, we consider the general half-odd-integer spin-$s$ cases. 
A similar analysis for the decoupled triangle with integer spin was 
also done in Ref.~\onlinecite{Charrier10}.
The Hamiltonian~\eqref{eq:3SpinHam} can be expressed as 
\begin{eqnarray}
H_{0,i} &=& \frac{J_\perp}{2} {\vec S}^2 + \frac{J'_\perp-J_\perp}{2} {\vec S}_{13}^2 
        - \frac{J_\perp}{2} {\vec s}_2^2 -\frac{J'_\perp}{2} \left( {\vec s}_1^2 + {\vec s}_3^2 \right) \nonumber \\
     &=& \frac{J_\perp}{2} S \left( S+1 \right) + \frac{J'_\perp - J_\perp}{2} S_{13} \left( S_{13}+1 \right) \nonumber \\
     & & - \frac{J_\perp+2J'_\perp}{2} s \left( s+1 \right),
\label{eq:3SpinHam2}
\end{eqnarray}
where we omitted the rung index $i$ for simplicity.
We also defined the total spin $\vec{S} = \vec{s}_1 + \vec{s}_2 + \vec{s}_3$
and  the bond spin ${\vec S}_{jk}={\vec s}_j+{\vec s}_k$,
and their magnitudes $S$ and $S_{jk}$ by 
$\vec{S}^2 = S(S+1)$ and ${\vec{S}_{jk}}^2 = S_{jk} (S_{jk}+1)$.
For the symmetric case ($J'_\perp = J_\perp$), the ground state 
belongs to the $S=1/2$ sector for any half-odd-integer spin-$s$. 
From the composition rule of angular momentum, $S_{13}$ obeys 
$S_{13} = s \pm 1/2$ and the ground state is four-fold degenerate 
with respect to $S_{13}$ and the eigenvalue $M= \pm 1/2$ of $S^z$. 
If we introduce a finite modulation of the rung couplings 
($J'_\perp \neq J_\perp$), degeneracy corresponding to $S_{13}$ is lifted. 
At some modulation $J'_\perp - J_\perp$, the energy level of 
the $S=1/2$ doublet ground state meets that of an $S=3/2$ quadruplet. 
For higher spin cases ($s > 1/2$), the ground state belongs to 
an $S=1/2$ doublet only in the restricted range of the modulation,
\begin{eqnarray}
\frac{2s}{2s+3} < \frac{J'_\perp}{J_\perp} < \frac{2 \left( s+1 \right)}{2s-1},
\label{eq:SRange}
\end{eqnarray}
centered around the symmetric point $J'_\perp/J_\perp = 1$. 
Since the bond spin $S_{13}$ is related to the parity for exchanging
between ${\vec s}_1$ and ${\vec s}_3$, we can derive the
effective Hamiltonian for $s>1/2$ by following the same
procedure as in the $s=1/2$ case.  
The detailed calculations are available in the Appendix.
As a result, the effective Hamiltonian for
general half-odd-integer spin-$s$ is given by
\begin{eqnarray}
H &=& \frac{J_\parallel}{3} \sum_{i=1}^L {\vec S}_i \cdot {\vec S}_{i+1} 
          \left[ 1 + 2\alpha \left( \tau^x_i \tau^x_{i+1} + \tau^y_i \tau^y_{i+1} \right) \right] \nonumber \\
   & &  + h \sum_{i=1}^L \tau^x_i,
\label{eq:SCHamG}
\end{eqnarray}
where 
\begin{eqnarray}
\alpha = \left( 2s+1 \right)^2, \hspace{10pt} h = \frac{2s+1}{2} \left( J'_\perp - J_\perp \right).
\label{eq.alpha_h}
\end{eqnarray}
This effective Hamiltonian for general half-odd integer $s$
has the same form as that~\eqref{eq:SCHam2} for $s=1/2$, 
and the $s$-dependence only enters through 
the coupling constants $\alpha$ and $h$.
However, it should be noted that this effective Hamiltonian is 
only valid in the range~\eqref{eq:SRange}.
This Hamiltonian coincides also with the one obtained for the three-leg
spin tube with a finite magnetization,~\cite{Plat12} 
if $h=0$ and ${\vec S}_i \cdot {\vec S}_{i+1}$ is taken as a constant.
The parameter $h$ corresponds to the strength of the asymmetry. 
For the symmetric case ($h=0$), earlier DMRG studies~\cite{Kawano97,Luscher04}
have confirmed that the ground state is in the dimer phase for
any value of $\alpha$.
Therefore, it is safe to say that the general half-odd-integer spin-$s$ 
tube exhibits the dimer phase at $h=0$ in the strong rung-coupling limit. 
This is consistent with the DMRG calculations~\cite{Nishimoto11a,Plat12} 
for the $s=3/2$ tube ($\alpha=16$) as well as those~\cite{Schnack04, Fouet06} 
on the $s=1/2$ `twisted' tube ($\alpha=2$).

%---[Preceding studies in the context of spin-orbital models]-----------------------

\subsection{Relationship with the spin-orbital model}

It is worth mentioning known models related to 
the effective Hamiltonian~\eqref{eq:SCHamG}.  
Indeed, it can be recognized to be a special form of 
the spin-orbital models which were first derived for 
the strong-coupling limit of the two-band Hubbard model 
at quarter-filling.~\cite{Kugel82}
If the parameters of the spin-orbital model are tuned 
in an appropriate way, we obtain the following Hamiltonian, 
\begin{eqnarray}
H &=& \sum_{i=1}^L \left[ J_1 {\vec S}_i \cdot {\vec S}_{i+1} + J_2 \left( T^x_i T^x_{i+1} + T^y_i T^y_{i+1} \right) \right. \nonumber \\
   & & \left. + K \left( {\vec S}_i \cdot {\vec S}_{i+1} \right) \left( T^x_i T^x_{i+1} + T^y_i T^y_{i+1} \right) \right], 
\label{eq:SOmodel}
\end{eqnarray}
where ${\vec T}_i$ is the orbital SU(2) matrix at site $i$. 
We observe that the model~\eqref{eq:SOmodel} reduces
to the effective Hamiltonian~\eqref{eq:SCHamG} with $h=0$,
if $J_1 = J_\parallel / 3$, $J_2 = 0$, and $K = 2 \alpha J_\parallel /3$. 
Hence, our effective Hamiltonian takes the same form as Eq.~\eqref{eq:SOmodel} 
by adding an extra chirality chain and turning off the transverse field. 
The system~\eqref{eq:SOmodel} has been studied 
by several groups.~\cite{Orignac00, Pati00, Kolezhuk01} 
At $J_1/K=3/4$ and $J_2/K=1/2$, it is exactly solvable
and the ground state is in a spontaneously dimerized phase where spin
and orbital form singlets alternatively along the chain.~\cite{Pati00, Kolezhuk01} 
As shown in the phase diagram of Ref.~\onlinecite{Pati00}, 
the dimer phase is extended to a wide parameter space for $J_1>0$, 
except for some $J_2 > J_1$ region where the ground state is 
composed of a spin ferromagnet and an orbital Fermi sea. 
Also, the possibility of the first order transition was
suggested in the $J_1>0, J_2 >0$ region; 
on the other hand, the dimer phase seems to be smoothly connected to 
the $J_2=0$ line, which corresponds to our effective 
Hamiltonian~\eqref{eq:SCHamG}, from the $J_1 >0, J_2 <0$ region.

%---[About the 3-leg spin tube with the FM leg and AFM rung]-------------------------------

We here mention a relation between the spin-orbital model~\eqref{eq:SOmodel} 
and the spin tube~\eqref{eq:3LegTubeHam} \emph{with ferromagnetic legs}.
For the symmetric spin tube with ferromagnetic legs, the corresponding 
effective Hamiltonian is again described by Eq.~\eqref{eq:SCHamG} 
with $J_\parallel <0$ and $h=0$. 
In Ref.~\onlinecite{Pati00}, this Hamiltonian has also been studied 
in the context of the spin-orbital model~\eqref{eq:SOmodel}, 
and its ground state is still in a dimer phase for $\alpha > 2$.
\footnote{In Ref.~\onlinecite{Pati00}, although they only considered Eq.~\eqref{eq:SOmodel} 
in the $K>0$ regime, a canonical transformation $T^{x,y}_i \rightarrow (-1)^i T^{x,y}_i$ 
makes a change $J_2 \rightarrow -J_2$ and $K \rightarrow -K$. 
Thus, their model still includes the effective
Hamiltonian~\eqref{eq:SCHamG} 
with $J_\parallel <0$ and $h=0$.} 
This result is consistent with a DMRG study in the $s=1/2$ spin tube 
with ferromagnetic legs, and a dimer phase is extended beyond 
the strong rung-coupling regime.~\cite{Nishimoto10}

%%%%%%%%%%%%%%%%%%%%%%%%%%%%%%%%%%%%%%%%%%%%%%%%%%%%%%%%%%%%%%%%%%%%%%%%%%%

\section{Field theoretical analysis}
\label{FieldTheory}

%---[Add a chirality chain to the strong-coupling Hamiltonian]-------------------

In this section, we analyze the effective
Hamiltonian~\eqref{eq:SCHamG} with a field theoretical method.
The bosonization approach combined with the RG method is a very
powerful and successful tool to study the low-energy properties of
1D systems.~\cite{GNT, Giamarchi} 
In this paper, Abelian bosonization is adopted. 
We first divide the effective Hamiltonian~\eqref{eq:SCHamG} into a single spin chain ($H_S$)
and interaction between the two degrees of freedom ($H_I$):
\begin{eqnarray}
H = H_S+H_I,
\label{eq:FullHam}
\end{eqnarray}
where
\begin{eqnarray*}
H_S &=& J_s \sum_{i=1}^L {\vec S}_i \cdot {\vec S}_{i+1}, \\
H_I &=& \lambda \sum_{i=1}^L {\vec S}_i \cdot {\vec S}_{i+1}
\left( \tau^x_i \tau^x_{i+1} + \tau^y_i \tau^y_{i+1} \right)
        + h \sum_{i=1}^L \tau^x_i. 
\end{eqnarray*}
Comparing (\ref{eq:FullHam}) with Eq.~\eqref{eq:SCHamG}, the couplings
constants are identified as $J_s=J_\parallel/3$ and
$\lambda=2J_\parallel \alpha /3$. 
The case of $h=0$ was studied in Ref.~\onlinecite{Orignac00}. 
We extend their bosonization approach in the presence of the
transverse field on chirality and derive a low-energy effective
theory which describes the phase transition in the spin
tube~\eqref{eq:3LegTubeHam}.

\subsection{Bosonization}

%---[Bosonization of the decoupled chain part]----------------------------

Let us start with the single spin chain $H_S$.
In the continuum limit, the bosonized Hamiltonian is given by
\begin{eqnarray}
H_S &\approx& \frac{v_s}{2\pi} \int dx \left[ K_s \left( \partial_x \theta_s \right)^2 
      + \frac{1}{K_s} \left( \partial_x \phi_s \right)^2 \right] \nonumber \\
& & -\frac{2 g_0}{\left( 2 \pi a_0 \right)^2} \int dx \cos \left( \sqrt{8} \phi_s \right), 
\label{eq:BosonHs}
\end{eqnarray}
where $x=a_0i$ ($a_0$: lattice spacing) and $g_0$ is a positive constant.
We here introduce the dual field for spin, satisfying the commutation relation,
\begin{eqnarray*}
\left[ \phi_s (x), \theta_s (x') \right] = -\frac{i \pi}{2} {\rm sgn} (x-x').
\end{eqnarray*}
For an isolated spin chain, the velocity $v_s$ and the Luttinger
parameter $K_s$ are exactly determined via Bethe
ansatz:~\cite{Giamarchi, Orignac00} $v_s=\pi J_sa_0/2$, $K_s=1$. 
Although the last term in Eq.~\eqref{eq:BosonHs} is marginally irrelevant 
and gives logarithmic corrections to various finite-size quantities, 
the low-energy properties of $H_S$ are described by 
a Tomonaga-Luttinger liquid (TLL) which is critical and gapless. 
However, if the interactions between spin and chirality are present, 
the $\cos (\sqrt{8} \phi_s)$ term plays a crucial role and could induce 
the phase transition.
Thus, we keep this term in the following analysis.

%---[Bosonization of the interaction part]-------------------------------------

Unlike the spin-orbital model~\eqref{eq:SOmodel}, our Hamiltonian~\eqref{eq:FullHam} 
originally does not contain a single chirality chain. 
However, as pointed out in Ref.~\onlinecite{Orignac00}, the interaction part $H_I$ can 
produce a finite velocity of chirality through the `mean-field-like' contribution. 
To see this, we split the nearest-neighbor spin correlation operator
into the expectation value with respect to the ground state of $H_S$,
$\left< {\vec S}_i \cdot {\vec S}_{i+1} \right>_0$, and the remaining
``normal-ordered'' part, $:{\vec S}_i \cdot {\vec S}_{i+1}:$, as
\begin{eqnarray}
{\vec S}_i \cdot {\vec S}_{i+1} = :{\vec S}_i \cdot {\vec S}_{i+1}: + \left< {\vec S}_i \cdot {\vec S}_{i+1} \right>_0. 
\label{eq:MFDecSpin}
\end{eqnarray}
Therefore, one can see that the interaction part $H_I$ gives a single chirality chain part, 
$-J_c \left( \tau^x_i \tau^x_{i+1} + \tau^y_i \tau^y_{i+1} \right)$, with a finite ferromagnetic exchange coupling 
$-J_c = \lambda \left< {\vec S}_i \cdot {\vec S}_{i+1} \right>_0 <0$. 
Like in a mean-field theory, chirality also gives a finite contribution to the spin chain through 
a similar decoupling as in Eq.~\eqref{eq:MFDecSpin}. 
However, considering $\lambda$ as a perturbation, the mean-field-like contribution 
merely provides small corrections to $v_s$ and $K_s$; 
furthermore, the SU(2) symmetry for spin ensures that $K_s$ stays at unity. 
Therefore, we neglect such corrections and suppose that they do not affect the qualitative properties 
of Eq.~\eqref{eq:FullHam} as long as we keep $\lambda$ as a perturbation. 

To obtain a bosonized form of $H_I$, we introduce the bosonization formulas 
for spin and chirality. 
In our bosonization language, spin operators are given by
\begin{eqnarray}
S^z_i \approx \frac{a_0}{\pi \sqrt{2}} \partial_x \phi_s + b (-1)^i \cos (\sqrt{2} \phi_s), 
\label{eq:BosonSz} \\
S^+_i \approx e^{i \sqrt{2} \theta_s} \left[ b (-1)^i + {\frac{1}{\pi}} \cos (\sqrt{2} \phi_s) \right], 
\label{eq:BosonSp}
\end{eqnarray}
where $b$ is a nonuniversal constant which can be numerically evaluated.~\cite{Lukyanov98, Hikihara98, Takayoshi10} 
Similarly, chirality operators are given by 
\begin{eqnarray}
\tau^+_i \approx \frac{e^{i \sqrt{2} \theta_c}}{\sqrt{2 \pi}} \left[ 1 + (-1)^i \cos (\sqrt{2} \phi_c) \right], 
\label{eq:BosonTp}
\end{eqnarray}
where $\phi_c$ and $\theta_c$ satisfy the commutation relation, 
\begin{eqnarray*}
\left[ \phi_c (x), \theta_c (x') \right] = -\frac{i \pi}{2} {\rm sgn} (x-x').
\end{eqnarray*}
They are obtained in a standard manner through the Jordan-Wigner
transformation.~\cite{Giamarchi} 
The staggered parts of the (normal-ordered) nearest-neighbor correlations 
are also given by
\begin{eqnarray}
(-1)^i : {\vec S}_i \cdot {\vec S}_{i+1} : \approx d \sin (\sqrt{2} \phi_s) 
\label{eq:BosonSS}
\end{eqnarray}
for spin, and 
\begin{eqnarray}
(-1)^i : \tau^x_i \tau^x_{i+1} + \tau^y_i \tau^y_{i+1} : \approx -\frac{1}{\pi} \sin (\sqrt{2} \phi_c) 
\label{eq:BosonTT}
\end{eqnarray}
for chirality.
The coefficient $d$ is a non-universal positive constant and 
has so far been obtained by several groups.~\cite{Takayoshi10, Orignac04} 
The staggered parts of chirality correlations~\eqref{eq:BosonTT} are 
also evaluated via the Jordan-Wigner transformation.~\cite{Orignac00, Orignac98} 
In the above formulas, we have neglected the higher-order terms because they only
give irrelevant perturbations to the low-energy physics in most cases.
Using Eqs.~\eqref{eq:BosonSz} to \eqref{eq:BosonTT}, the perturbation
$H_I$ can be expressed in terms of boson fields as
\begin{eqnarray}
H_I &\approx& \frac{v_c}{2\pi} \int dx 
\left[ K_c (\partial_x \theta_c)^2 + \frac{1}{K_c} (\partial_x \phi_c)^2 \right] \nonumber \\
&& -\frac{\lambda d}{\pi a_0} \int dx \sin (\sqrt{2} \phi_s) \sin (\sqrt{2} \phi_c) \nonumber \\ 
&& +\frac{h}{\sqrt{2\pi} a_0} \int dx \cos (\sqrt{2} \theta_c),
\end{eqnarray}
where $v_c=J_c a_0$ and $K_c=2$. 

%---[Full bosonized Hamiltonian]----------------------------------------

Finally, we obtain the bosonized form of the full Hamiltonian~\eqref{eq:FullHam}, 
\begin{eqnarray}
H 
&\approx& \sum_{\nu=s,c} \frac{v_\nu}{2\pi} \int dx \left[ K_\nu \left( \partial_x \theta_\nu \right)^2 
   + \frac{1}{K_\nu} \left( \partial_x \phi_\nu \right)^2 \right] \nonumber \\
& & - \frac{2g_0}{(2 \pi a_0)^2} \int dx \cos (\sqrt{8} \phi_s) \nonumber \\
& & - \frac{2g_1}{(2 \pi a_0)^2} \int dx \sin (\sqrt{2} \phi_s) \sin (\sqrt{2} \phi_c) \nonumber \\ 
& & + \frac{2g_2}{(2 \pi a_0)^2} \int dx \cos (\sqrt{2} \theta_c), 
\label{eq:BosonFullHam}
\end{eqnarray}
where the coupling constants are given by 
\begin{eqnarray}
& g_1=2\pi a_0 \lambda d, \hspace{10pt} & g_2=\sqrt{2 \pi^3} a_0 h. 
\end{eqnarray}
In this Hamiltonian, both $g_1$ and $g_2$ are relevant since their scaling dimensions are given by 
$(K_s+K_c)/2 \sim 3/2$ and $1/(2K_c) \sim 1/4$, respectively. 
To investigate the low-energy physics, we perform 
the perturbative RG analysis in the following section. 

%---[Renormalization group equations]-------------------------------------

\subsection{Renormalization group equations}

Using the operator product expansion,~\cite{Cardy} we obtain the following RG equations, 
\begin{subequations}
\label{eq:RGEqKc}
\begin{eqnarray}
\frac{dy_0}{dl} &=& \left( 2-K_s \right) y_0 -\frac{1}{8} y_1^2, \\
\frac{dy_1}{dl} &=& \left( 2-\frac{K_s+K_c}{2} \right) y_1 +\frac{1}{2} y_0 y_1, \\
\frac{dy_2}{dl} &=& \left( 2-\frac{1}{2K_c} \right) y_2, \\
\frac{dK_s}{dl} &=& -\frac{1}{2} K_s^2 y_0^2 -\frac{1}{16} K_s^2 y_1^2, \\
\frac{dK_c}{dl} &=& -\frac{1}{16} K_c^2 y_1^2 + \frac{1}{8} y_2^2,
\end{eqnarray}
\end{subequations}
where $l$ is the logarithmic length scale.
In these equations, the velocity difference between spin and chirality chains is neglected, 
i.e., $u \equiv v_s = v_c$, and we set $ y_i = g_i /(\pi u)$, for simplicity.

%---[Expected phases without a transverse field]---------------------------

Let us first consider the symmetric tube ($g_2=0$), which has been
studied with the RG method in Ref.~\onlinecite{Orignac00}. 
In this case, only $g_1$ is relevant and the ground-state energy is 
minimized when the condition
\begin{eqnarray}
\sin (\sqrt{2} \phi_s) \sin (\sqrt{2} \phi_c) =1
\end{eqnarray}
is satisfied. 
Then, the field for spin (chirality) is locked to $\phi_s = \pi/\sqrt{8}$ ($\phi_c = \pi/\sqrt{8}$) 
or $\phi_s = -\pi/\sqrt{8}$ ($\phi_c = -\pi/\sqrt{8}$), and the dimer order parameters 
for both spin and chirality have finite expectation values 
from Eqs.~\eqref{eq:BosonSS} and \eqref{eq:BosonTT}. 
This indicates the presence of a long-range dimer order, in agreement with the
previous DMRG calculations.~\cite{Kawano97, Luscher04} 
In this phase, the spin and chirality form singlets on bonds alternatively and
translational symmetry is spontaneously broken.  
Consequently, both spin and chirality excitations are gapped. 

%---[Expected phases with a transverse field]-------------------------------

Next we introduce the asymmetry of the rung triangle ($g_2 \neq 0$). 
In the limit of $|g_2| \to \infty$, the chirality degrees of freedom are 
completely frozen out and only spin degrees of freedom remain; 
therefore, the system is in a TLL phase. 
The coupling $g_2$ is highly relevant and it would be expected that 
the chirality is polarized along the $x$-direction as soon as $g_2$ is switched on.
However, the other coupling $g_1$ competes with $g_2$ because $g_1$
pins the field $\phi_c$ whereas $g_2$ pins its dual field
$\theta_c$. Thus, a finite critical value of $g_{2c}$ to bring the
system from the dimer-ordered state to another one may exist.
We will discuss this possibility in more detail in the next subsection.

%---[Derivation of the effective Hamiltonian under the strong transverse field]----------------

\subsection{Effective theory for the spin degrees of freedom}

In this subsection, we derive an effective theory only for the spin
degrees of freedom, which describes the transition between the dimer and TLL phases, 
in the case of sufficiently strong $g_2$.
Here, we apply the two-cutoff scaling method.~\cite{GNT, EHKim99}
The coupling $g_2$, the initial value of which is $g_2(l=0) \propto h$,
is the most relevant perturbation at the weak-coupling
fixed point (two-component TLL of spin and chirality).
Thus we assume that, while $g_2$ grows to $O(1)$,
other couplings remain perturbatively small under the RG.
We define the logarithmic length scale $\bar{l}$ such
that $g_2(l=\bar{l}) = O(1)$.
At this length scale, the field $\theta_c$ is locked to 
the potential minimum; $\theta_c$ and its dual
$\phi_c$ can be safely integrated out as massive fields.

%---[Perturbative expansion]--------------------------------------------------

The partition function is written as 
\begin{eqnarray}
{\mathcal Z} = \int {\mathcal D} \phi_s {\mathcal D} \phi_c e^{-S_0 [\phi_s] -S_c [\phi_c] -S_1 [\phi_s, \phi_c]}, 
\end{eqnarray}
where
\begin{eqnarray}
S_0 [\phi_s] &=& \frac{u}{2\pi {\bar K}_s} \int d^2r \left[ \frac{1}{u^2} (\partial_\tau \phi_s)^2 + (\partial_x \phi_s)^2 \right], \\
S_c [\phi_c] &=& \frac{u}{2\pi {\bar K}_c} \int d^2r \left[ \frac{1}{u^2} (\partial_\tau \phi_c)^2 + (\partial_x \phi_c)^2 \right] \nonumber \\
                & & + \frac{2{\bar g}_2}{(2\pi a_0)^2} \int d^2 r \cos (\sqrt{8} \theta_c), \\
S_1 [\phi_s, \phi_c] &=& \frac{2}{(2\pi a_0)^2} \int d^2 r \left[ -{{\bar g}_0} \cos (\sqrt{8} \phi_s) \right. \nonumber \\ 
                          & & \left. + {\bar g}_1 \sin (\sqrt{2} \phi_s) \sin (\sqrt{2} \phi_c) \right], 
\end{eqnarray}
and ${\bar X}=X(l={\bar l})$ denotes a renormalized coupling and $r \equiv (\tau, x)$. 
We expand the partition function up to second order in $S_1$, 
\begin{eqnarray}
{\mathcal Z} = z_c \int {\mathcal D} \phi_s e^{-S_0 [\phi_s]} \left[1 - \left< S_1 \right>_c + \frac{1}{2} \left< S_1^2 \right>_c + \cdots \right], 
\label{eq:ExpandZ}
\end{eqnarray} 
where the expectation value is defined by 
\begin{eqnarray}
\left< \cdots \right>_c = \frac{1}{z_c} \int {\mathcal D} \phi_c (\cdots) e^{-S_c [\phi_c]}, 
\end{eqnarray}
and $z_c=\int {\mathcal D} \phi_c e^{-S_c [\phi_c]}$ is a constant. 
The first order contribution is easily evaluated as
\begin{eqnarray}
\left< S_1 \right>_c = -\frac{2{{\bar g}_0}}{(2\pi a_0)^2} \int d^2 r \cos (\sqrt{8} \phi_s). 
\end{eqnarray}
A non-vanishing term in the second order contribution is given by
\begin{eqnarray}
\left< S_1^2 \right>_c &=& \frac{4{\bar g}_1^2}{(2\pi a_0)^4} \int d^2 r_1 d^2 r_2 \nonumber \\ 
                            & & \times \sin \left( \sqrt{2} \phi_s (r_1) \right) \sin \left( \sqrt{2} \phi_s (r_2) \right) \nonumber \\ 
                            & & \times \left< \sin \left( \sqrt{2} \phi_c (r_1) \right) \sin \left( \sqrt{2} \phi_c (r_2) \right) \right>_c. 
\label{eq:av_S1}
\end{eqnarray}
Since the field $\theta_c$ is ordered in the present case, 
its dual field $\phi_c$ is disordered and the correlation function of 
$\phi_c$ decays exponentially. 
Thus, we consider only the short-range contribution 
from $r_1 \sim r_2$ and approximate as follows:
\begin{eqnarray}
\left< \sin \left( \sqrt{2} \phi_c (r_1) \right) \sin \left( \sqrt{2} \phi_c (r_2) \right) \right>_c \approx D \delta (r_1-r_2), 
\end{eqnarray}
where $D$ is a positive nonuniversal constant. 
Then, Eq.~\eqref{eq:av_S1} is rewritten as
\begin{eqnarray}
\left< S_1^2 \right>_c \approx -\frac{2 D {\bar g}_1^2}{(2 \pi a_0)^2} \int d^2r \cos (\sqrt{8} \phi_s). 
\end{eqnarray}
Further, we reexponentiate the expectation values in Eq.~(\ref{eq:ExpandZ}) 
and obtain the effective action, 
\begin{eqnarray}
S_{\rm eff} [\phi_s] = S_0 [\phi_s] - \frac{2{\bar g}_0 - D {\bar g}_1^2}{(2\pi a_0)^2} \int d^2 r \cos (\sqrt{8} \phi_s), 
\end{eqnarray}
where ${\mathcal Z} \approx z_c \int {\mathcal D} \phi_s e^{-S_{\rm eff} [\phi_s]}$. 

Finally, the effective spin Hamiltonian above the scale ${\bar l}$ is 
given by the usual sine-Gordon model, 
\begin{eqnarray}
H_{\rm eff} &=& \frac{u}{2\pi} \int dx \left[ {\bar K}_s (\partial_x \theta_s)^2 + \frac{1}{{\bar K}_s} (\partial_x \phi_s)^2 \right] \nonumber \\
          & & - \frac{2G}{(2 \pi a_0)^2} \int dx \cos (\sqrt{8} \phi_s), 
\label{eq:EffHamSpin}
\end{eqnarray}
where $G={{\bar g}_0}-D {\bar g}_1^2/2$. 

Although this Hamiltonian does not contain explicitly the transverse field $h$
in the original model, it does depend on $h$.
This is because the length scale $\bar{l}$ and thus the renormalized couplings
at $l=\bar{l}$ are determined by $g_2(l=0) \propto h$.
In the leading approximation,
$\bar{l} \propto - \frac{1}{2 - 1/(2K_c)} \log{h} \sim - \frac{4}{7} \log{h}$, 
$\bar{g}_0 \sim g_0$, and
$\bar{g}_1 \sim g_1 \exp{\left[(2- \frac{K_s+K_c}{2})\bar{l}\right]} \sim g_1 h^{-2/7}$.
Inclusion of subleading terms in the RG equation changes these
values, although the qualitative picture remains the same
as long as $\bar{g}_0$ and $\bar{g}_1$ are small.
On the other hand, thanks to the SU(2) symmetry, 
${\bar K}_s$ should be always 1. 
The sine-Gordon interaction in the last term of 
Eq.~\eqref{eq:EffHamSpin} has the scaling dimension $2$ and thus is marginal. 
In fact, it is marginally relevant or irrelevant, depending on the sign of $G$. 
If $h$ is sufficiently strong, ${\bar g}_1$ does not change much 
from its initial value $g_1 (l=0)$ and $G$ stays positive; 
the system is in a TLL phase with a marginally irrelevant term. 
If $h$ decreases, $g_1$ grows under the RG flow and $G$ can be negative. 
In this case, $G$ is marginally relevant and $\phi_s$ is
locked into one of its minima $\pm \pi/\sqrt{8}$; 
the system goes into a dimer phase. 
Therefore, at a critical value of $h$ where the
marginal coupling $G$ exactly vanishes, we can expect a phase
transition between the TLL and dimer phases.  
This belongs to a particular variant of BKT transition 
with the SU(2) symmetry in the entire region, as known 
in the $s=1/2$ $J_1$-$J_2$ chain.~\cite{Affleck89, Okamoto92}
We check these expectations numerically in the following section.

%%%%%%%%%%%%%%%%%%%%%%%%%%%%%%%%%%%%%%%%%%%%%%%%%%%%%%%%%%%%%%%%%%%%%%%%%%%%%%%

\section{Numerical analysis}
\label{Numerical}

In this section, we attempt to confirm numerically
the above picture obtained from the analytical arguments.
First we verify the phase transition in the effective
Hamiltonian~\eqref{eq:SCHamG} numerically, based on
the level spectroscopy method with exact diagonalization.
Furthermore, we also look for the phase transition
in the original spin tube model by numerical
estimation of central charge with DMRG.
Hereafter, we focus on the $s=1/2$ case and set $J_\parallel=1$ 
as energy unit unless otherwise stated.

%---[Level spectroscopy analyses]----------------------------------

\subsection{Level Spectroscopy for the effective Hamiltonian}
\label{sec.levelspec}

%###########################################################
\begin{figure}
\includegraphics[clip,width=8cm]{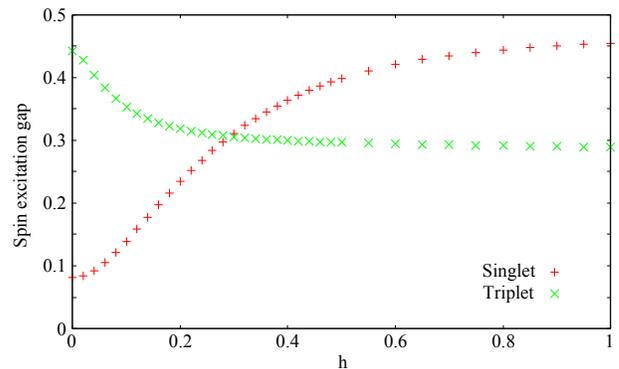}
\caption{(Color Online) Excitation spectra for the model~\eqref{eq:SCHamG} 
with $s=1/2$, $L=16$, as functions of $h$. 
Symbols $+$ and $\times$ denote the singlet-singlet and 
singlet-triplet excitation gaps, respectively.}
\label{fig:ExcitationGap}
\end{figure}
%###########################################################

%---[Level crossing in energy spectra]------------------------------

Here we apply the level spectroscopy method~\cite{Okamoto92} to
the effective Hamiltonian~\eqref{eq:SCHamG}.
This enables us to detect the BKT transition between the dimer and TLL phases. 
Finite-size systems of $L=8$, $10$, $12$, $14$, and $16$ are
studied using exact diagonalization technique with the PBC.  
In the context of the sine-Gordon theory, the gaps of singlet-singlet and
singlet-triplet excitations are expected to cross at the BKT
transition point for a finite-size system. 
As an example, the results of the gaps for $L=16$ are shown 
in FIG.~\ref{fig:ExcitationGap}. 
We can clearly see the crossing at $h \simeq 0.24$.

%---[Scaling dimensions]------------------------------------------

\begin{figure}
\includegraphics[clip,width=8cm]{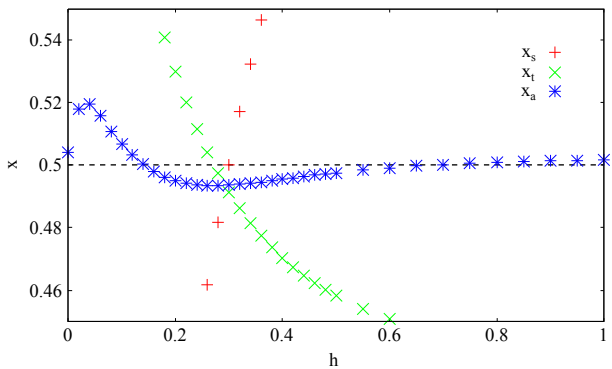}
\caption{(Color Online) Scaling dimensions fo the model~\eqref{eq:SCHamG} 
with $s=1/2$, $L=16$, as functions of $h$. 
Symbols $+$, $\times$, and $+ \hspace{-7.2pt} \times$ denote 
$x_s$, $x_t$, and $x_a$, respectively (see text).}
\label{fig:ScalingDimension}
\end{figure}

In order to check the universality class of the present BKT
transition, we estimate the scaling dimensions. 
By conformal field theory (CFT), the scaling dimensions are related 
to the excitation spectra of a finite-size system as follows,~\cite{Cardy84}
\begin{eqnarray}
E_n (L) - E_{gs} (L) = \frac{2\pi v}{L} x_n (L),
\end{eqnarray}
where $E_n (L)$ and $E_{gs} (L)$ are energies of an excited state and
of the ground state, respectively, for a system with length $L$, $v$
is the spin-wave velocity, and $x_n (L)$ is the scaling dimension.
The subscript $n$ is an index labeling the type of excitations (see below). 
In the TLL phase, the scaling dimensions for the first singlet
($n=s$) and triplet ($n=t$) excitations have logarithmic corrections
from the marginal perturbation:~\cite{Affleck89}
\begin{eqnarray}
x_s (L) = \frac{1}{2} + \frac{3}{4} \frac{1}{\log L} + {\mathcal O} \left( \frac{1}{L^2} \right), \\
x_t (L) = \frac{1}{2} - \frac{1}{4} \frac{1}{\log L} + {\mathcal O} \left( \frac{1}{L^2} \right). 
\end{eqnarray}
These logarithmic corrections are expected to vanish
only at the BKT transition point. 
Away from the transition point, they quickly become 
significant and often cause severe problems to scaling analyses. 
It is thus very useful to remove the corrections by taking
an average $x_{a} = (x_s+3x_t)/4$. \cite{Ziman87}
To obtain the scaling dimensions, we still need an estimate of the spin-wave velocity $v$.
It is typically determined by
\begin{eqnarray}
v (L) = \frac{L}{2\pi} \left[ E^{k} (L) - E_{gs} (L) \right],
\end{eqnarray}
for a system with length $L$, where $E^k (L)$ is an energy of the
first-excited state with wave vector $k=2\pi/L$ 
(the ground state is characterized by $k=0$).  
We calculate $v (L)$ for $L=8 - 16$, and then extrapolate it 
to the thermodynamic limit via a fitting function,
\begin{eqnarray}
v(L) = v + A_0 \frac{1}{L^2} + A_1 \frac{1}{L^4} + \cdots,
\end{eqnarray}
where $A_0$ and $A_1$ are fitting parameters. 
For all values of $h(<1)$, the fitting works perfectly 
despite neglecting the higher-order terms. 
The estimated values of $x_s$, $x_t$, and $x_a$ as
functions of $h$ are shown in FIG.~\ref{fig:ScalingDimension}.  
The original scaling dimensions $x_s$ and $x_t$ are near $1/2$ 
only around the crossing point $h_{\rm cross}$, 
while $x_a$ remains close to $1/2$ at $h \gtrsim h_{\rm cross}$.  
This indicates that the crossing point corresponds to the BKT transition 
between the dimer and TLL phases. 
Note that $h_{\rm cross}$ is equivalent to the crossing point
between the singlet-singlet and singlet-triplet gaps shown in FIG.~\ref{fig:ExcitationGap}.  
The similar results have also been obtained for the original $s=1/2$ 
spin tube~\eqref{eq:3LegTubeHam}.~\cite{Sakai08}

%##################################################################
%###################################################################

%###################################################################
\begin{figure}
\includegraphics[clip,width=8cm]{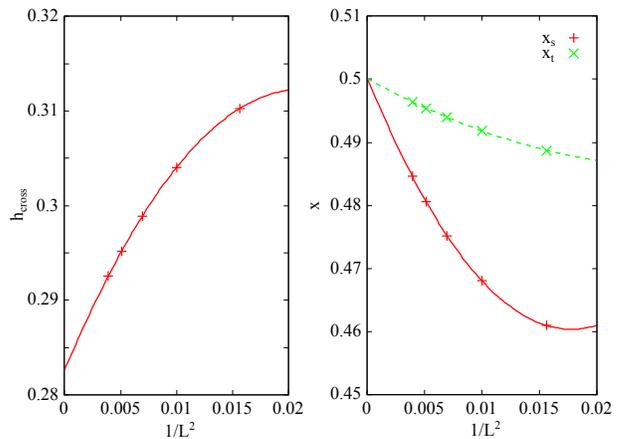}
\caption{(Color Online) Size scaling of the crossing point between 
$x_s$ and $x_t$ (left panel), and of the scaling dimensions 
at the transition point $h_c=0.283$ (right panel), as functions of $1/L^2$. 
In the right panel, symbols $+$ and $\times$ denote $x_s$ and $x_t$, respectively. 
Fitting functions~\eqref{eq:FitCP} and \eqref{eq:FitSD} are used 
for the finite-size scaling.}
\label{fig:CPandSD}
\end{figure}
%###################################################################

%---[BKT transition point]-------------------------------------------------

At the BKT transition, there is no logarithmic correction, 
but other corrections attributed to the $x=4$ irrelevant 
perturbation exist.~\cite{Cardy86a, Cardy86b} 
Therefore, we extrapolate the crossing point and scaling dimensions 
to the thermodynamic limit, using the following scaling functions,
\begin{eqnarray}
h_{\rm cross} (L) &=& h_c + B_0 \frac{1}{L^2} + B_1 \frac{1}{L^4} + \cdots, 
\label{eq:FitCP} \\
x (L) &=& x_c + C_0 \frac{1}{L^2} + C_1 \frac{1}{L^4} + \cdots, 
\label{eq:FitSD}
\end{eqnarray}
where $B$'s and $C$'s are fitting parameters. 
The scaling analyses are shown in FIG.~\ref{fig:CPandSD}. 
The crossing point in the thermodynamic limit is evaluated as
\begin{equation}
h_c=0.283,
\end{equation}
which corresponds to 
the critical value of $h$ between the dimer and TLL phases.
We note that the critical point $h_c$ is obtained with respect
to the leg coupling $J_\parallel = 1$.
It remains finite in the strong rung-coupling limit
$J'_\perp, J_\perp \to \infty$.
Since $h = J'_\perp - J_\perp$, the critical asymmetry \emph{ratio} 
$(J'_{\perp,c} - J_\perp)/J_\perp$ is small (proportional to $1/J_\perp$) 
in the strong rung-coupling limit $J_\perp \rightarrow \infty$. 

In Ref.~\onlinecite{Sakai08} and \onlinecite{Sakai05}, Sakai \textit{et al.} studied 
the original spin-$1/2$ tube by the level spectroscopy method using exact diagonalization 
and found the dimer phase in the region $0.95 \lesssim J'_\perp/J_\perp \lesssim 1.05$ 
for $J_\parallel/J_\perp=0.2$. 
Although the critical asymmetry in our study, $|J'_\perp-J_\perp|/J_\parallel = 0.283$, 
is obtained in the strong-rung coupling limit, this is considerably close to the value 
$|J'_\perp-J_\perp|/J_\parallel \sim 0.25$ obtained by Sakai \textit{et al}. 
Thus the present work strongly supports that a finite but small critical asymmetry exists 
in the strong rung-coupling region. 

At $h=h_c$, both the singlet and triplet scaling dimensions are
extrapolated to $x_c=0.500$ in the thermodynamic limit.  
This is consistent with the theoretical prediction, 
giving a strong evidence supporting the theory developed in Sec.~\ref{FieldTheory}. 
Our analysis in this section is limited to the value of $\alpha$ in
Eq.~\eqref{eq.alpha_h} corresponding to $s=1/2$. 
Nevertheless, from Eq.~\eqref{eq:SCHamG}, it is easily speculated that 
essentially the same BKT transition occurs for the general-$s$ 
($s \ge 1/2$) cases, i.e., for any $\alpha$, in the strong rung-coupling regime 
since $\alpha$ changes only the strength of coupling between 
the spin and chirality chains.

%---[DMRG analyses on the 3-leg spin tube]--------------------------------

\subsection{DMRG for the three-leg spin tube}

%---[Entanglement entropy and central charge]------------------------------

In order to confirm the validity of the analysis based
on the effective Hamiltonian, here we investigate
the transition point in the original spin tube~\eqref{eq:3LegTubeHam} 
by calculating the central charge using the DMRG method.~\cite{White92, Schollwock05} 

The central charge, which can be evaluated 
via the von-Neumann entanglement entropy, gives a direct 
information to determine the universality classes of 1D quantum systems. 
From CFT, the von-Neumann entanglement entropy of a subsystem 
with length $l$ in a periodic system 
with length $L$ has the following scaling form,~\cite{Calabrese04} 
\begin{eqnarray}
S_L (l) = \frac{c}{3} \ln \left[ \frac{L}{\pi} \sin \left( \frac{\pi l}{L} \right) \right] + c', 
\end{eqnarray}
for a critical phase, where $c$ is the central charge and $c'$ is a nonuniversal constant. 
Using this formula, we estimate $c$ as~\cite{Nishimoto11b} 
\begin{eqnarray}
c (L) = \frac{3 \left[ S_L \left( \frac{L}{2}-1 \right) - S_L \left( \frac{L}{2} \right) \right]}{\ln \left[ \cos \left( \frac{\pi}{L} \right) \right]}. 
\label{eq:c_from_SL}
\end{eqnarray}
In the thermodynamic limit, the central charge is $c=1$ for the gapless TLL phase 
while $c=0$ for the gapped dimer phase. 

%##############################################################
\begin{figure}
\includegraphics[clip,width=8cm]{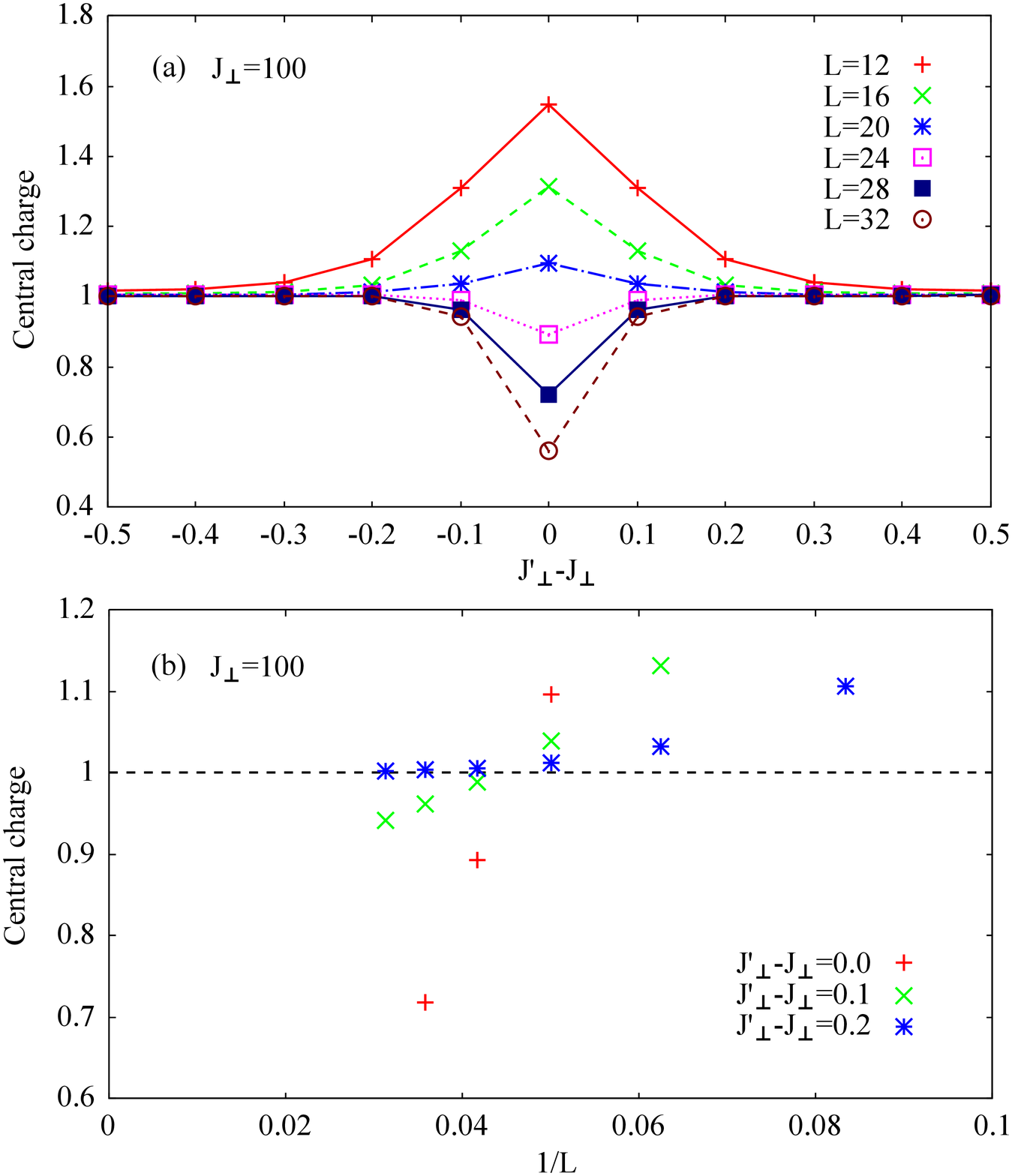}
\caption{(Color Online) (a) Central charge of the $s=1/2$ three-leg spin 
tube~\eqref{eq:3LegTubeHam} with $J_\perp=100$, as functions of $J'_\perp-J_\perp$. 
(b) System-size dependence of the central charge near the symmetric limit, 
as a function of $1/L$.}
\label{fig:CentralChargeDMRGJ100}
\end{figure}
%##############################################################

%##############################################################
\begin{figure}
\includegraphics[clip,width=8cm]{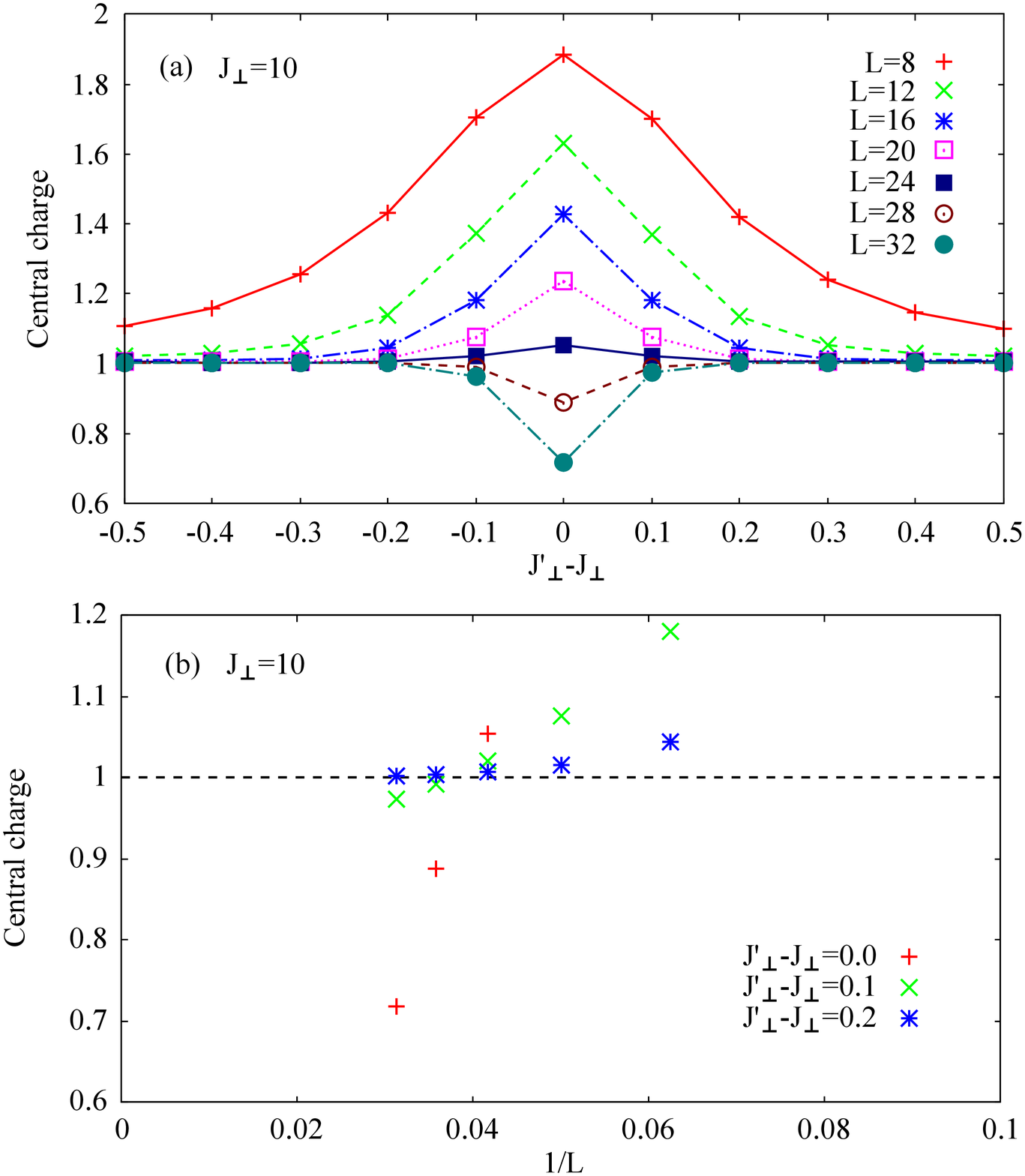}
\caption{(Color Online) (a) Central charge of the $s=1/2$ three-leg spin 
tube ~\eqref{eq:3LegTubeHam} with $J_\perp=10$, as functions of $J'_\perp-J_\perp$. 
(b) System-size dependence of the central charge near the symmetric limit, 
as a function of $1/L$.}
\label{fig:CentralChargeDMRGJ10}
\end{figure}
%##############################################################

%---[BKT transition point]---------------------------------------------

We study systems with $L=8$, $12$, $16$, $20$, $24$, $28$, and $32$
under the PBC, and keep $m \simeq 800$, $1400$, $2200$, $3000$,
$4000$, $5000$, and $6000$ density-matrix eigenstates, respectively. 
An evaluation of the central charge with Eq.~\eqref{eq:c_from_SL} 
needs extremely high accuracy of numerical data but gives a
sharp insight for the thermodynamic limit even with small systems. 
The results for two strong rung-coupling
cases $J_\perp=100$ and $J_\perp=10$
are shown in FIG.~\ref{fig:CentralChargeDMRGJ100} and
\ref{fig:CentralChargeDMRGJ10}, respectively.  
At the symmetric point ($J'_\perp=J_\perp$), the central charge deviates 
toward $c(L) \simeq 2$ for small $L$ because the system behaves 
as two decoupled spin and chirality chains from the viewpoint of RG; 
however apparently, it decreases with increasing $L$, cuts across $c=1$, 
and goes down to zero as approaching the thermodynamic limit. 
The similar behaviors are seen at $|J'_\perp-J_\perp|=0.1$ for 
both of $J_\perp=100$ and $J_\perp=10$. 
The decay of $c(L)$ with decreasing $1/L$ is much slower
than that at the symmetric point. 
However, it is natural to expect that it is extrapolated to zero 
in the thermodynamic limit once it comes down to $c<1$, 
since the existence of any nontrivial fixed point with $0<c<1$
is not expected. 
Therefore, the system is still in the dimer phase at $|J'_\perp-J_\perp|=0.1$. 
On the other hand, $c(L)$ appears to converge to $1$ with
decreasing $1/L$ at $|J'_\perp-J_\perp|=0.2$.
This means that the system behaves as a TLL at $|J'_\perp-J_\perp|=0.2$,
within the system sizes up to $L=32$ studied here.
However, it should be noted that this does not exclude the possibility
that the system is actually gapped in the thermodynamic limit
but the correlation length is larger than $32$.
Indeed, the exponentially large correlation length is expected
in the gapped side of the BKT transition. Thus the present
result does not contradict with the level spectroscopy analysis of
the effective Hamiltonian in Sec.~\ref{sec.levelspec},
that implies the critical point at $|J'_\perp-J_\perp| \sim 0.283$.

Although a precise estimation of the BKT transition point is difficult
in the present direct analysis due to the exponentially
large correlation
length and restriction of the system lengths, we can conclude that the
dimer phase is extended to at least $|J'_\perp-J_\perp| \leq 0.1$.
This is indeed consistent with the level spectroscopy analysis
of the effective Hamiltonian.
As expected from the above theoretical and numerical analyses for the
effective Hamiltonian~\eqref{eq:SCHamG}, the BKT transition
between the TLL and dimer phases occurs at a very small value of
$|J'_\perp-J_\perp|$ relative to $J_\perp$, in the
strong rung-coupling regime.

%%%%%%%%%%%%%%%%%%%%%%%%%%%%%%%%%%%%%%%%%%%%%%%%%%%%%%%%%%%%%%%%

\section{Conclusion}
\label{Conclusion}

%---[Consequences from bosonization]--------------------------------------

In this paper, we have discussed the quantum phase transitions of the
three-leg spin tube induced by the modulation of the rung couplings.
In the strong rung-coupling limit $J_\perp \gg J_\parallel$,
the $s=$ half-odd-integer tube is mapped
onto an effective model which is regarded as a special type of the
spin-orbital models, where the orbital degrees of freedom correspond
to the chirality ones in the original spin tube.
The effective model was
studied using the bosonization and RG techniques.  
For the symmetric tube ($J'_\perp = J_\perp$), 
we found that the system is in the dimer phase with a finite excitation gap 
for the general spin-$s$ cases. 
Then, we investigated an effect of the asymmetric modulation of 
the rung couplings, which appears as a transverse chirality field 
in the effective Hamiltonian.  
For the asymmetric tube ($J'_\perp \neq J_\perp$), we confirmed that the
phase transition between the dimer and TLL phases is a BKT transition
governed by the ordinary sine-Gordon model.

%---[Consequences from numerical results]-------------------------------------

Furthermore, we performed the level spectroscopy analysis of the
strong rung-coupling effective Hamiltonian using
exact diagonalization and
determine the BKT transition point to be $|J'_\perp - J_\perp|=0.283$.
Moreover, using the DMRG method we calculated the central charge of the
original $s=1/2$ spin tube in the strong rung-coupling
regime $J_\perp=10$ and $100$.
Although the direct numerical analysis of the original spin tube model
does not allow a precise determination of the transition point,
we found that the dimer phase is robust against the modulation up to
$|J'_\perp-J_\perp|=0.1$, in agreement with the analysis based
on the effective Hamiltonian.
Thus, we conclude that the dimer phase
persists for very small but a finite modulation of the rung couplings
in the strong rung-coupling regime.  It was also shown that the universal
properties of the three-leg spin tube are essentially the same as
those of the frustrated $J_1$-$J_2$ chain, as suggested in
Ref.~\onlinecite{Sakai08} and \onlinecite{Sakai10}.

%---[Remining problems]---------------------------------------------

Our analyses are based on the strong rung-coupling
expansion for $J_\perp, J'_\perp \gg J_\parallel$ and 
it is quite hard to speculate the low-energy properties 
in the intermediate and weak rung-coupling regimes.
However, very recently, it was suggested that the BKT transition for
the rung modulation occurs even in the weak rung-coupling regime with using
the bosonization approach.~\cite{Furuya13}

%%%%%%%%%%%%%%%%%%%%%%%%%%%%%%%%%%%%%%%%%%%%%%%%%%%%%%%%%%%%%

\begin{acknowledgments}
We thank S. C. Furuya, M. Nakamura, K. Okamoto, and M. Tsuchiizu for
fruitful discussions.  YF was supported in part by Global COE Program
``the Physical Science Frontier'' and Leading Graduate School for
Frontiers of Mathematical Science and Physics (FMSP),
while MO was supported in part by JSPS KAKENHI Grants Nos. 21540381
and 25400392.
MO also thanks the hospitality of Laboratoire de Physique Th\'{e}orique,
IRSAMC, Universit\'e Paul Sabatier and CNRS, Toulouse,
where a part of this work was carried out.
\end{acknowledgments}

%%%%%%%%%%%%%%%%%%%%%%%%%%%%%%%%%%%%%%%%%%%%%%%%%%%%%%%%%%%%%%

\appendix*
\section{Strong rung-coupling Hamiltonian for general spin-\texorpdfstring{$s$}{s}}
\label{GeneralS}

In this appendix, we give a detailed derivation of the strong rung-coupling 
Hamiltonian~\eqref{eq:SCHamG} for general spin-$s$. 
For the angular momentum algebraic technique, see Ref.~\onlinecite{Edmonds}. 

%---[Construction of the 3-site ground state for general-S]----------------------

We start with the construction of eigenstates with total spin $S$. 
For this purpose, ${\vec s}_1$ and ${\vec s}_3$ are first combined, 
and then ${\vec s}_2$ is added to ${\vec S}_{13}$:
\begin{eqnarray}
\left| S_{13}; S M \right>
&=& \left| \left( \left( s_3 s_1 \right) S_{13} s_2 \right) S M \right> \nonumber \\
&=& \sum_{m_1 m_2 m_3} \left( s m_3 s m_1 | S_{13} \ m_1+m_3 \right) \nonumber \\
& & \times \left( S_{13} \ m_1+m_3 \ s m_2 | S M \right) \left| m_1 m_2 m_3 \right>, \nonumber \\
\label{eq:StateS13}
\end{eqnarray}
where the summation of $m_j$ runs over from $-s$ to $s$, 
and $\left( s m s' m' | S M \right)$ is the Clebsch-Gordan coefficient. 
By applying the exchange operator ${\hat P}$ 
between ${\vec s}_1$ and ${\vec s}_3$ to Eq.~\eqref{eq:StateS13}, we obtain
\begin{eqnarray}
{\hat P} \left| S_{13}; S M \right> 
&=& \sum_{m_1 m_2 m_3} \left( s m_3 s m_1 | S_{13} \ m_1+m_3 \right) \nonumber \\ 
& & \times \left( S_{13} \ m_1+m_3 \ s m_2 | S M \right) \left| m_3 m_2 m_1 \right> \nonumber \\
&=& \left( -1 \right)^{2s-S_{13}} \left| S_{13}; S M \right>,
\end{eqnarray}
where $\left( s m s' m' | S M \right) = \left( -1 \right)^{s+s'-S} \left( s' m' s m | S M \right)$ is used. 
Thus, $\left| S_{13}; S M \right>$ is an eigenstate of ${\hat P}$. 
Here, we restrict the $S=1/2$ case, i.e., $S_{13}= s \pm 1/2$, 
and then the eigenstates of ${\hat P}$ are denoted as 
\begin{eqnarray}
\left| M + \right> \equiv \left| S_{13}=s+\frac{1}{2}; S=\frac{1}{2}, M \right> 
\label{eq:StatePM1}
\end{eqnarray}
for $P= \left( -1 \right) ^{s-1/2}$ and 
\begin{eqnarray}
\left| M - \right> \equiv \left| S_{13}=s-\frac{1}{2}; S=\frac{1}{2}, M \right> 
\label{eq:StatePM2}
\end{eqnarray}
for $P= \left( -1 \right) ^{s+1/2}$. 
These states are also eigenstates of the Hamiltonian~\eqref{eq:3SpinHam} and 
their energy eigenvalues are given by Eq.~\eqref{eq:3SpinHam2}. 

%---[Alternative expression of the 3-site ground state]-------------------------------

For later convenience, we introduce an alternative expression of the eigenstate with $S$. 
Here, just as Eq.(\ref{eq:StateS13}) was obtained, we first combine ${\vec s}_2$ 
with ${\vec s}_3$ and then add ${\vec s}_1$ to ${\vec S}_{23}$. 
As a result, we have
\begin{eqnarray}
& & \left| \left( s_1 \left( s_2 s_3 \right) S_{23} \right) S M \right> \nonumber \\ 
&=& \sum_{m_1 m_2 m_3} \left( s m_2 s m_3 | S_{23} \ m_2+m_3 \right) \nonumber \\
& & \times \left( s m_1 S_{23} \ m_2+m_3 | S M \right) \left| m_1 m_2 m_3 \right>, 
\end{eqnarray}
and this expression is related to Eq.~\eqref{eq:StateS13} by
\begin{eqnarray}
& & \left< \left( s_1 \left( s_2 s_3 \right) S_{23} \right) S M | S_{13}; S M \right> \nonumber \\ 
&=& \left( -1 \right) ^{3s+S_{23}-S_{13}+S} \sqrt{ \left( 2S_{13}+1 \right) \left( 2S_{23}+1 \right) } 
\left\{ \begin{array}{ccc} s & s & S_{23} \\ s & S & S_{13} \end{array} \right\}, \nonumber \\ 
\end{eqnarray}
where $\left\{ \begin{array}{ccc} s_1 & s_2 & S_{12} \\ s_3 & S & S_{13} \end{array} \right\}$ is the Wigner's 6-j symbol. 
Therefore, we can write
\begin{eqnarray}
& & \left| S_{13}; S M \right> \nonumber \\ 
&=& \sum_{S_{23}} \left( -1 \right) ^{3s+S_{23}-S_{13}+S} \sqrt{ \left( 2S_{13}+1 \right) \left( 2S_{23}+1 \right) } \nonumber \\ 
& & \times \left\{ \begin{array}{ccc} s & s & S_{23} \\ s & S & S_{13} \end{array} \right\} 
\sum_{m_1 m_2 m_3} \left( s m_2 s m_3 | S_{23} \ m_2+m_3 \right) \nonumber \\ 
& & \times \left( s m_1 S_{23} \ m_2+m_3 | S M \right) \left| m_1 m_2 m_3 \right>.
\label{eq:StateS132} 
\end{eqnarray}

%---[Perturbation theory]----------------------------------------

Now, our aim is to derive the strong rung-coupling Hamiltonian for $H=H_0+H_1$, 
where $H_0$ is the unperturbed part, 
\begin{eqnarray}
H_0 = \sum_{i=1}^L \left[ J_\perp \left( {\vec s}_{i,1} \cdot {\vec s}_{i,2} + {\vec s}_{i,2} \cdot {\vec s}_{i,3} \right)
         + J'_\perp {\vec s}_{i,1} \cdot {\vec s}_{i,3} \right], \nonumber \\
\label{eq:H0}
\end{eqnarray}
and $H_1$ is the perturbation,
\begin{eqnarray}
H_1 = J_\parallel \sum_{i=1}^L \sum_{j=1}^3 {\vec s}_{i+1,j} \cdot {\vec s}_{i,j}.
\label{eq:H1}
\end{eqnarray}
The projection operator onto the Hilbert space consisting of direct products of 
the $S=1/2$ states is defined as 
\begin{eqnarray}
Q = \prod_{i=1}^L Q_i, \hspace{10pt} Q_i = \sum_{M_i P_i} \left| M_i P_i \right> \left< M_i P_i \right|,
\end{eqnarray}
where $P_i$ takes the values $\pm 1$ corresponding to Eqs.~\eqref{eq:StatePM1} and \eqref{eq:StatePM2}. 
The strong rung-coupling Hamiltonian is then written as $H_{SC} = QHQ$. 
Using Eqs.~\eqref{eq:TauZ} and \eqref{eq:3SpinHam2}, 
the unperturbed part $QH_0Q$ is easily calculated as 
\begin{eqnarray}
Q H_0 Q = \frac{2s+1}{2} \left( J'_\perp - J_\perp \right) \sum_{i=1}^L \tau^z_i + {\rm const.}
\label{eq:QH0Q}
\end{eqnarray}
In order to calculate the perturbation part $QH_1Q$, we consider the projected representation 
of $s_{i,j}^\mu$ ($\mu=x,y,z$), 
\begin{eqnarray}
Q_i s_{i,j}^\mu Q_i 
&=& \sum_{M_i M'_i P_i P'_i} \left| M_i P_i \right> \left< M_i P_i \right| 
s_{i,j}^\mu \left| M'_i P'_i \right> \left< M'_i P'_i \right| \nonumber \\ 
&=& \sum_{M_i M'_i P_i P'_i} \zeta^{(j)}_{P_i P'_i} \left| M_i \right> \left< M_i \right| 
S_i^\mu \left| M'_i \right> \left< M'_i \right|, \nonumber \\
\end{eqnarray}
where we define
\begin{eqnarray}
\zeta^{(j)}_{PP'} 
& \equiv & \frac{ \left< M P \right| s^\mu_j \left| M' P' \right> }{ \left< M \right| S^\mu \left| M' \right> } \nonumber \\ 
&=& \frac{ \left< M=\uparrow P || s_j || M'=\uparrow P' \right> }
              { \left< M=\uparrow || S || M'=\uparrow \right> } \nonumber \\
&=& 2 \left< M=\uparrow P \right| s^z_j \left| M'=\uparrow P' \right>
\end{eqnarray}
and use the Wigner-Eckart theorem. 
The perturbation part is now expressed as 
\begin{eqnarray}
Q H_1 Q 
&=& J_\parallel \sum_{i=1}^L {\vec S}_i \cdot {\vec S}_{i+1} \sum_{P_i P_{i+1} P'_i P'_{i+1}} \nonumber \\ 
& & \times \sum_{j=1}^3 \zeta^{(j)}_{P_i P'_i} \zeta^{(j)}_{P_{i+1} P'_{i+1}} 
\left| P_i P_{i+1} \right> \left< P'_i P'_{i+1} \right|. \nonumber \\
\label{eq:QH1Q}
\end{eqnarray}
Using the parity operator ${\vec \tau}$ and the Hermiticity of the coefficient 
$\zeta^{(j)}_{PP'} = \left( \zeta^{(j)}_{P'P} \right)^*$, 
Eq.~\eqref{eq:QH1Q} is rewritten as 
\begin{widetext}
\begin{eqnarray}
Q H_1 Q 
&=& J_\parallel \sum_{i=1}^L {\vec S}_i \cdot {\vec S}_{i+1} \sum_{j=1}^3
       [\frac{1}{2}(\zeta_{++}^{(j)}+\zeta_{--}^{(j)}) + (\zeta_{++}^{(j)}-\zeta_{--}^{(j)})\tau_i^z + 2\zeta_{+-}^{(j)}\tau_i^x] \nonumber \\ 
 & & \times [\frac{1}{2}(\zeta_{++}^{(j)}+\zeta_{--}^{(j)}) + (\zeta_{++}^{(j)}-\zeta_{--}^{(j)})\tau_{i+1}^z + 2\zeta_{+-}^{(j)}\tau_{i+1}^x]. 
\label{eq:QH1Q2}
\end{eqnarray}
From the exchange symmetry (${\vec s}_1 \leftrightarrow {\vec s}_3 $), 
the coefficient $\zeta^{(j)}_{PP'}$ obeys
\begin{eqnarray}
\zeta^{(1)}_{PP'} = PP' \zeta^{(3)}_{PP'}, \hspace{10pt} \zeta^{(2)}_{PP'} = \delta_{PP'} \zeta^{(2)}_{PP}. 
\end{eqnarray}
Here, the only thing we have to do is to evaluate the coefficient $\zeta^{(j)}_{PP'}$. 
The calculation of $\zeta^{(2)}_{PP}$ may be relatively easy: It is given by
\begin{eqnarray}
\zeta^{(2)}_{++} 
&=& 2 \left< \uparrow + \right| s^z_2 \left| \uparrow + \right> \nonumber \\
&=& 2 \sum_{m_1 m_2 m_3} m_2 \left( s m_3 s m_1 | s+\frac{1}{2} \ m_1+m_3 \right)^2 
      \left( s+\frac{1}{2} \ m_1+m_3 \ s m_2 | \frac{1}{2} \frac{1}{2} \right)^2 \nonumber \\ 
&=& 2 \sum_{m_2=-s}^s m_2 \left( s+\frac{1}{2} \ \frac{1}{2}-m_2 \ s m_2 | \frac{1}{2} \frac{1}{2} \right)^2 \nonumber \\ 
&=& -\frac{2s}{3}, 
\end{eqnarray}
and
\begin{eqnarray}
\zeta^{(2)}_{--} 
&=& 2 \left< \uparrow - | s^z_2 | \uparrow - \right> \nonumber \\ 
&=& 2 \sum_{m_1 m_2 m_3} m_2 \left( s m_3 s m_1 | s-\frac{1}{2} \ m_1+m_3 \right)^2  
      \left( s-\frac{1}{2} \ m_1+m_3 \ s m_2 | \frac{1}{2} \frac{1}{2} \right)^2 \nonumber \\ 
&=& 2 \sum_{m_2=-s+1}^s m_2 \left( s-\frac{1}{2} \ \frac{1}{2}-m_2 \ s m_2 | \frac{1}{2} \frac{1}{2} \right)^2 \nonumber \\ 
&=& \frac{2 \left( s+1 \right)}{3}, 
\end{eqnarray}
where we use the unitarity of the Clebsch-Gordan coefficient, and the following formulas, 
\begin{eqnarray}
\left( s+\frac{1}{2} \ \frac{1}{2}-m \ s m | \frac{1}{2} \frac{1}{2} \right)^2 
&=& \frac{s-m+1}{(s+1)(2s+1)}, \nonumber \\ 
\left( s-\frac{1}{2} \ \frac{1}{2}-m \ s m | \frac{1}{2} \frac{1}{2} \right)^2 
&=& \frac{s+m}{s(2s+1)}.
\label{eq:CGFormula}
\end{eqnarray}
It seems to be slightly more complicated to calculate $\zeta^{(1)}_{PP'}$: 
Using the expression in Eq.~\eqref{eq:StateS132}, we obtain 
\begin{eqnarray}
\zeta^{(1)}_{++} 
&=& 2 \left< \uparrow + \right| s^z_1 \left| \uparrow + \right> \nonumber \\ 
&=& 8 \left( s+1 \right)^2 
\left\{ \begin{array}{ccc} s & s & s+\frac{1}{2} \\ s & \frac{1}{2} & s+\frac{1}{2} \end{array} \right\}^2 
\sum_{m_1=-s}^s m_1 \left( s+\frac{1}{2} \ \frac{1}{2}-m_1 \ s m_1 | \frac{1}{2} \frac{1}{2} \right)^2 \nonumber \\
& & + 8s \left( s+1 \right)
\left\{ \begin{array}{ccc} s & s & s-\frac{1}{2} \\ s & \frac{1}{2} & s+\frac{1}{2} \end{array} \right\}^2 
\sum_{m_1=-s+1}^s m_1 \left( s-\frac{1}{2} \ \frac{1}{2}-m_1 \ s m_1 | \frac{1}{2} \frac{1}{2} \right)^2 \nonumber \\ 
&=& \frac{2s+3}{6},
\end{eqnarray}
\begin{eqnarray}
\zeta^{(1)}_{--}
&=& 2 \left< \uparrow - \right| s^z_1 \left| \uparrow - \right> \nonumber \\ 
&=& 8s \left( s+1 \right)
\left\{ \begin{array}{ccc} s & s & s+\frac{1}{2} \\ s & \frac{1}{2} & s-\frac{1}{2} \end{array} \right\}^2 
\sum_{m_1=-s}^s \left( s+\frac{1}{2} \ \frac{1}{2}-m_1 s m_1 | \frac{1}{2} \frac{1}{2} \right)^2 \nonumber \\ 
& & + 8s^2
\left\{ \begin{array}{ccc} s & s & s-\frac{1}{2} \\ s & \frac{1}{2} & s-\frac{1}{2} \end{array} \right\}^2 
\sum_{m_1=-s+1}^s m_1 \left( s-\frac{1}{2} \ \frac{1}{2}-m_1 \ s m_1 | \frac{1}{2} \frac{1}{2} \right)^2 \nonumber \\ 
&=& \frac{-2s+1}{6},
\end{eqnarray}
and
\begin{eqnarray}
\zeta^{(1)}_{+-} 
&=& 2 \left< \uparrow + \right| s^z_1 \left| \uparrow - \right> \nonumber \\ 
&=& -8 \sqrt{s(s+1)^3} 
\left\{ \begin{array}{ccc} s & s & s+\frac{1}{2} \\ s & \frac{1}{2} & s+\frac{1}{2} \end{array} \right\} 
\left\{ \begin{array}{ccc} s & s & s+\frac{1}{2} \\ s & \frac{1}{2} & s-\frac{1}{2} \end{array} \right\} 
\sum_{m_1=-s}^s m_1 \left( s+\frac{1}{2} \ \frac{1}{2}-m_1 \ s m_1 | \frac{1}{2} \frac{1}{2} \right)^2 \nonumber \\ 
& & -8 \sqrt{s^3(s+1)}
\left\{ \begin{array}{ccc} s & s & s-\frac{1}{2} \\ s & \frac{1}{2} & s+\frac{1}{2} \end{array} \right\} 
\left\{ \begin{array}{ccc} s & s & s-\frac{1}{2} \\ s & \frac{1}{2} & s-\frac{1}{2} \end{array} \right\} 
\sum_{m_1=-s+1}^s m_1 \left( s-\frac{1}{2} \ \frac{1}{2}-m_1 \ s m_1 | \frac{1}{2} \frac{1}{2} \right)^2 \nonumber \\ 
&=& \frac{2s+1}{2 \sqrt{3}},
\end{eqnarray}
\end{widetext}
where we have used Eq.~\eqref{eq:CGFormula} and the following 6-j symbol formulas,
\begin{eqnarray}
\left\{ \begin{array}{ccc} s & s & s+\frac{1}{2} \\ s & \frac{1}{2} & s+\frac{1}{2} \end{array} \right\}
&=& (-1)^{3s+1/2} \frac{1}{4(s+1)}, \nonumber \\ 
\left\{ \begin{array}{ccc} s & s & s-\frac{1}{2} \\ s & \frac{1}{2} & s-\frac{1}{2} \end{array} \right\} 
&=& (-1)^{3s-1/2} \frac{1}{4s}, \nonumber \\ 
\left\{ \begin{array}{ccc} s & s & s+\frac{1}{2} \\ s & \frac{1}{2} & s-\frac{1}{2} \end{array} \right\}
&=& \left\{ \begin{array}{ccc} s & s & s-\frac{1}{2} \\ s & \frac{1}{2} & s+\frac{1}{2} \end{array} \right\} \nonumber \\ 
&=& (-1)^{3s+1/2} \frac{1}{4} \sqrt{ \frac{3}{s(s+1)} }. 
\end{eqnarray}
Substituting the above expressions of $\zeta^{(j)}_{PP'}$ into Eq.~\eqref{eq:QH1Q2}, 
the perturbation part is obtained as 
\begin{eqnarray}
Q H_1 Q 
&=& \frac{J_\parallel}{3} \sum_{i=1}^L {\vec S}_i \cdot {\vec S}_{i+1} \nonumber \\ 
& & \times \left[ 1 +2(2s+1)^2 \left( \tau^x_i \tau^x_{i+1} + \tau^z_i \tau^z_{i+1} \right) \right] \nonumber \\ 
\label{eq:QH1Q3}
\end{eqnarray}
Incorporating Eqs.~\eqref{eq:QH0Q} and \eqref{eq:QH1Q3} and 
applying an appropriate unitary transformation, 
we finally obtain the strong rung-coupling Hamiltonian \eqref{eq:SCHamG}. 

\bibliography{References}

\end{document}